\documentclass[twocolumn]{jpsj2} 

\title{Doping Dependence of Two Energy Scales \\in the Tunneling Spectra of Superconducting La$_{2-x}$Sr$_{x}$CuO$_{4}$}
\author{Takuya \textsc{Kato}\thanks{E-mail address: tkato@rs.kagu.tus.ac.jp}, Toshiyuki \textsc{Maruyama}, Shigeki \textsc{Okitsu}, and Hideaki \textsc{Sakata}}
\inst{Department of Physics, Tokyo University of Science,\\1-3 Kagurazaka, Shinjuku-ku, Tokyo 162-8601, Japan}

\abst{
Scanning tunneling microscopy and spectroscopy have been performed on superconducting La$_{2-x}$Sr$_{x}$CuO$_{4}$ ($x=0.06, 0.12, 0.16$, and $0.21$) at 4.2~K.
Atomically resolved surface topography has been observed on the $ab$ surface, which exhibits a one-dimensional structure corresponding to the buckling pattern of the orthorhombic phase.
The tunneling spectra of all samples exhibit kinks at about $\pm 5$~meV in addition to the energy gap, indicating the existence of two energy scales.
While the gap magnitude varies on the length scale of a few nm and steadily increases upon underdoping, the kink energy is spatially uniform on the surface and almost independent of the doping level.
}
\kword{high-$T_{\mathrm{c}}$ superconductor, gap inhomogeneity, La$_{2-x}$Sr$_{x}$CuO$_{4}$ (LSCO), STM/STS}
\begin{document}
\maketitle

\section{Introduction}
Clarification of the nature of the superconducting gap is crucial for understanding the mechanism of high-temperature superconductors (HTSs) because it will provide much information about the pairing of electrons.
The energy gap in the electronic density of states has been observed by various spectroscopic techniques.~\cite{two-gaps_review_2007}
Scanning tunneling spectroscopy (STS) using a scanning tunneling microscope (STM), which directly probes the quasiparticle local density of states (LDOS) in real space with high energy resolution, is one of the most powerful techniques among them.
Since the discovery of HTSs, many STS experiments have been performed to observe the LDOS in HTSs, and anomalous behaviors of the energy gap have been revealed:~\cite{sts_review_2007}
(i) the energy gap $\Delta$ is not uniform in real space and varies on the length scale of a few nm,~\cite{inhomo_pan_2001, inhomo_lang_2002, inhomo_mcelroy_prl_2005, inhomo_mcelroy_science_2005,kato_prb} 
(ii) the spatial average of $\Delta$ increases with decreasing hole concentration in spite of the suppression of the superconducting critical temperature $T_\mathrm{c}$,
and (iii) an energy gap called {\it pseudogap} appears below the temperature $T^{*} > T_\mathrm{c}$, and $\Delta$ in the superconducting state smoothly evolves into the pseudogap across $T_\mathrm{c}$.~\cite{2212_pseudogap_renner_1998, 2201_pseudogap_kugler_2001}
Since $T^{*}$ increases with decreasing hole concentration, $\Delta$ seems to be in agreement with $T^{*}$.
Thus, another energy scale according with $T_{\mathrm{c}}$, which is probably intrinsic to high-$T_{\mathrm{c}}$ superconductivity, has been anticipated.

Recently we discovered the existence of another lower energy scale in optimally doped Bi$_{2}$Sr$_{2-x}$La$_{x}$CuO$_{6+\delta}$ (La-Bi2201).~\cite{2201_machida_2006}
The lower energy scale has been observed as a {\it kink} structure inside the energy gap in the tunneling spectra.
Similar low-energy kinks have been observed in Bi$_{2}$Sr$_{2}$CaCu$_{2}$O$_{8+\delta}$ (Bi2212).~\cite{inhomo_mcelroy_prl_2005,2212_two-gaps_gomes_2007} 
Thus, it should be clarified whether the lower energy scale exists in HTSs other than Bi-based cuprates and to investigate how its energy depends on doping or $T_{\mathrm{c}}$.
In comparison with Bi-based cuprates, because of their poor cleavage, La-based cuprates have been less explored by STM/STS, although they have been extensively studied by various methods other than STM/STS.
The restriction of STM/STS studies to cuprates having good cleavage has made it difficult to understand the universal nature of the electronic states of HTSs.
Our previous STM/STS study on optimally doped La$_{2-x}$Sr$_{x}$CuO$_{4}$ (LSCO) demonstrated the applicability of STM/STS to LSCO.~\cite{kato_prb}
Although atomic resolution was not achieved, the existence of the gap inhomogeneity in LSCO as well as in Bi2212 was elucidated.
In that study, we did not observe the kink structure in the tunneling spectra because of insufficient energy resolution.
In this paper, we report on a STS study of the superconducting LSCO from the underdoping regime to the overdoping regime with high enough energy resolution to investigate the low-energy structure.
Atomically resolved surface topography has been successfully observed on the $ab$ surface, and the tunneling spectra of all samples clearly show the kinks inside the energy gap in the tunneling spectra, indicating the existence of two energy scales in LSCO as well as in Bi-based cuprates.
While the gap magnitude shows nanoscale inhomogeneity and strongly depends on the doping level, the kink energy is spatially uniform and almost independent of the doping level.

\section{Experimental Procedure}
Single crystals of LSCO ($x=0.06$, $0.12$, $0.16$, and $0.21$) used in this study were grown by the traveling-solvent floating-zone technique.~\cite{tsfz_lsco_kojima_1989}
It was confirmed that the difference between the actual and nominal Sr contents was within $\pm 0.004$ by inductively coupled plasma mass spectrometry and electron-probe microscope analysis. 
The superconducting transition temperature $T_{\mathrm{c}}$, defined as the onset temperature of the transition observed by magnetization measurements, was 14, 30, 38, and 27~K for $x = 0.06$, $0.12$, $0.16$, and $0.21$, respectively. 
The $a_{\mathrm{HTT}}$ axes (Cu--O--Cu bonding directions) and $c$ axis of the grown crystals were determined by the X-ray Laue method at room temperature (in the high-temperature tetragonal (HTT) phase).
The STM/STS experiments were performed using a laboratory-built low-temperature STM/STS system at 4.2~K. 
The samples were fractured \textit{in situ} to expose the $ab$ surface.
The bias voltage $V$ was applied to the sample, namely, negative and positive bias correspond to occupied and unoccupied states, respectively.
Tunneling spectra d$I$/d$V(V)$ were acquired by numerical differentiation of the measured current-voltage characteristics. 
The STS measurements were carried out by measuring the tunneling spectra at a regularly spaced grid of $64 \times 64$ or $128 \times 128$ points over a scan area of $40 \times 40$~nm$^2$. 
The experimental details have been described elsewhere.~\cite{vortex_ynbc_sakata_2000,kato_prb}

\section{Results and Discussion}

\subsection{Surface topography}

\begin{figure}[b]
\begin{center}
\includegraphics[width=75mm]{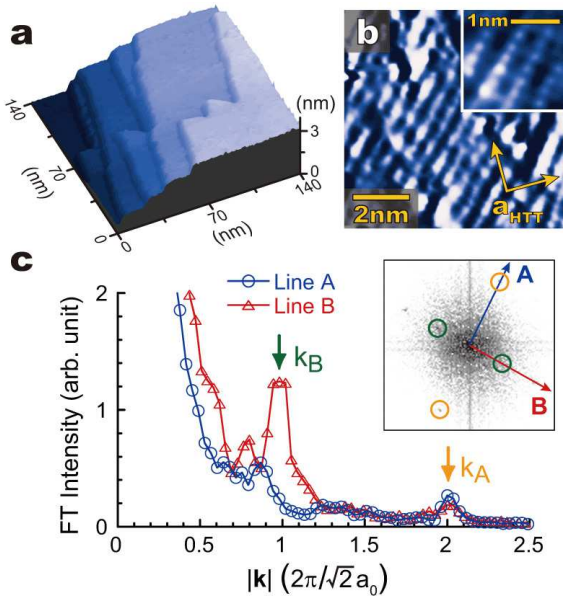}
\end{center}
\caption{\label{Topography}
(a) Large-scale topography of a fractured $ab$ surface of LSCO showing six terraces.
(b) High-resolution image of a terrace.
(main) $I=30$~pA, $V=-1.7$~V and (inset) $I=30$~pA, $V=-1.2$~V.
(c) (inset) Fourier transform (FT) of an atomically resolved image. 
The primitive FT peaks along arrows A and B are circled in orange and green, respectively.
(main) Line profiles of the FT image along arrows A and B depicted in the inset.
The positions of the FT peaks are indicated by the vertical arrows in the respective colors of the circles in the inset.
}
\end{figure}

Before exhibiting the spectroscopic results, we show the surface topography of LSCO in Fig.~\ref{Topography}, because almost nothing is known about the surface structure of LSCO.
As shown in Fig.~\ref{Topography}(a), the exposed $ab$ surface exhibits atomically flat terraces having a step height of $\sim c_{0}/2$ ($c_{0}$ is the lattice constant along the $c$ axis) attributed to the layered structure of LSCO along the $c$ axis.~\cite{kato_prb}
Some of the terraces have an area exceeding $50 \times 50$~nm$^{2}$, which is enough to perform STS to measure the spatial distribution of the LDOS.
By magnifying a terrace, we can see a one-dimensional structure with a period of $\sim 5.4$~\AA\ regardless of the Sr content of the sample, as shown in Fig.~\ref{Topography}(b). 
The direction of the bright lines was rotated by 45$^{\circ}$ with respect to the $a_{\mathrm{HTT}}$ axes.
As shown in the inset of Fig.~\ref{Topography}(b), we can see that the bright spots are arranged along the lines with a period of $\sim 2.7$~\AA.
These configurations are clearly exhibited in the Fourier transform (FT) image shown in Fig.~\ref{Topography}(c).
Primitive FT peaks corresponding to a period of $\sim 2.7$~\AA\ appear at $|\vec{k}_{\mathrm{A}}| = 2 \times (2\pi/\sqrt{2}a_{0})$ along one of the $(110)_{\mathrm{HTT}}$ directions (line A), where $a_{0} \sim 3.8$~\AA\ denotes the Cu--O--Cu distance.
Along another $(110)_{\mathrm{HTT}}$ direction (line B), they appear at half of $|\vec{k}_{\mathrm{A}}|$ ($|\vec{k}_{\mathrm{B}}| = 2\pi/\sqrt{2}a_{0}$).
Hence, the observed atomic configuration is a 2.7~\AA $\times$ 5.4~\AA\ rectangular lattice aligned in the $(110)_{\mathrm{HTT}}$ direction.
The appearance of this surface structure was independent of the sample bias voltage, and thus it probably indicates a crystalline structure.

We conclude that this one-dimensional structure originates from the buckling of the CuO$_2$ plane in the low-temperature orthorhombic (LTO) phase for following reasons:
(i) all the samples used in this study are in the LTO phase at 4.2~K,~\cite{lsco_xrd_1993, lsco_xrd_1994}
(ii) the direction of the one-dimensional structure is consistent with the tilting direction of the CuO$_6$ octahedra, 
and (iii) the period of $\sim 5.4$~\AA\ is consistent with the lattice constant in the LTO phase.
In LSCO, the CuO$_2$ and La(Sr)O layers are possibly exposed by the fracture.
Since the observed 2.7~\AA $\times$ 5.4~\AA\ surface unit cell agrees well with the configuration of the upward in-plane oxygen atoms of the tilted CuO$_6$ octahedra, it is most probable that the topmost layer is the CuO$_2$ layer and the observed bright spots are the in-plane oxygen atoms.
On the other hand, the one-dimensional modulation is also expected in the La(Sr)O layer because the tilt of the CuO$_{6}$ octahedra results in the shift of the apical oxygen atoms, as observed in manganese oxides.~\cite{bcmo_renner_nature_2002}
To understand the surface structure of LSCO and its STM topograph completely, precise theoretical or computational work is needed.~\cite{2212_topo_model_bansil_2007}
In any case, since the La(Sr)O layer is thought to be insulating, the low-energy LDOS measured by STS---the scope of the present paper and discussed in the next section---is largely contributed by the electronic states of the CuO$_2$ plane.

\subsection{Existence and doping dependence of two energy scales observed in the tunneling spectra}

\begin{figure*}[tb]
\begin{center}
\includegraphics[width=170mm]{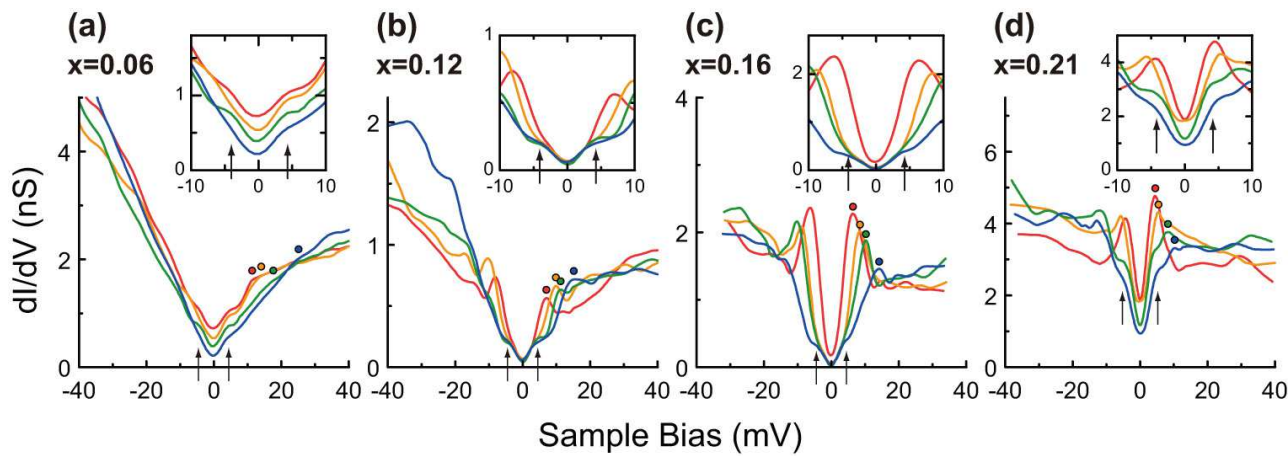}
\end{center}
\caption{\label{SpectrumEvolution}
Doping dependence of tunneling spectra.
Spectra having various gap energies $\Delta_1$ (indicated by filled circles in respective colors) for each doping level are shown.
All spectra exhibit kinks at $\Delta_2 \sim \pm 5$~meV (upper arrows) independent of $\Delta_1$ and the doping level.
Insets show enlarged low-energy parts of the same spectra.
}
\end{figure*}

To investigate how the tunneling spectra evolve with doping, STS was performed on a terrace in each sample.
The representative spectra are shown in Fig.~\ref{SpectrumEvolution}.
As the hole concentration is reduced from the overdoping regime ($x=0.21$), the background of the spectra evolves from a flat (metallic) one to a V-shaped (semiconducting) one.
In addition to this change, the asymmetry between positive and negative biases increases.
These tendencies are consistent with those observed in the other cuprate HTSs and represent the doping evolution of the background electronic structure intrinsic to the doped Mott insulator.~\cite{cncoc_hangauri_nature_2004,asymmetry_randeria_2005,asymmetry_anderson_2006,asymmetry_kohsaka_2007}

\begin{figure}[tb]
\begin{center}
\includegraphics[width=75mm]{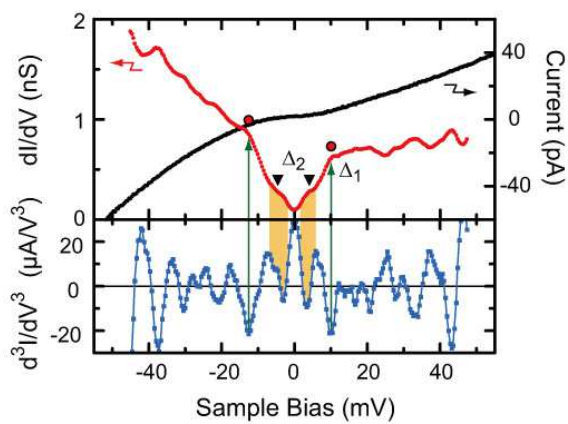}
\end{center}
\caption{\label{RepresentativeSpectrum}
Representative tunneling spectrum showing two energy scales in LSCO ($x=0.12$).
Raw data ($I(V)$, black), the tunneling spectrum ($\mathrm{d}I/\mathrm{d}V(V)$, red), and its second derivative ($\mathrm{d}^3I/\mathrm{d}V^3(V)$, blue) are shown.
The edges of the energy gap $\Delta_1$ (indicated by two filled circles) are represented by the negative peaks in the $\mathrm{d}^3I/\mathrm{d}V^3(V)$ curve (green arrows).
The curve of $\mathrm{d}^3I/\mathrm{d}V^3(V)$ also reveals the existence of the lower characteristic energy scale $\Delta_2$ (indicated by two inverted triangles) as the dips below $\Delta_1$ (orange regions).
} 
\end{figure}

\begin{figure}[tb]
\begin{center}
\includegraphics[width=75mm]{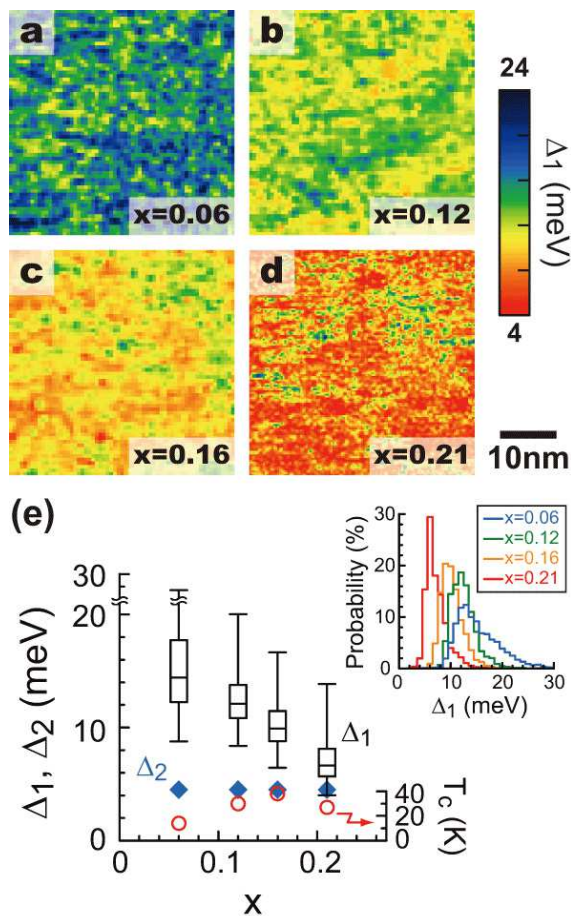}
\end{center}
\caption{\label{GapDistribution}
(a)-(d) Gap maps for $x = 0.06$, $0.12$, $0.16$, and $0.21$.
The color and length scales are identical for all maps.
(e) Doping dependences of $\Delta_1$, $\Delta_2$, and $T_{\mathrm{c}}$.
The distribution of $\Delta_1$ is shown using a box-and-whisker plot.~\cite{box_whisker}
The inset shows the histograms of $\Delta_1$.
}
\end{figure}

An energy gap, whose edge at the positive bias is indicated by each filled circle in Fig.~\ref{SpectrumEvolution}, was observed in the background electronic structure.
Its energy spatially varies on the length scale of a few nm, and is known as the nanoscale electronic inhomogeneity (see Figs.~\ref{GapDistribution}(a)-\ref{GapDistribution}(d)).~\cite{kato_prb}
Hereafter, we refer to the energy of this gap as $\Delta_1$.
Although the spectra in the overdoped and optimally doped samples exhibit rather sharp gap edge peaks, the peaks were rarely observed in the underdoped samples, particularly in the spectra having large $\Delta_1$.
To define the value of $\Delta_1$ in such a broad spectral profile, we use the second derivative of the tunneling spectra ($\mathrm{d}^3I/\mathrm{d}V^3(V)$).
In Fig.~\ref{RepresentativeSpectrum}, a representative tunneling spectrum of the underdoped sample without gap edge peaks and its second derivative are shown.
As indicated by the green arrows in the figure, the $\mathrm{d}^3I/\mathrm{d}V^3(V)$ curve shows two negative peaks near the gap edges, corresponding to the maxima of negative curvature.
We adopt the positions of these peaks to represent the gap edges.
Due to the asymmetric V-shaped background of the spectrum, the gap edge at the negative bias in the underdoped sample becomes extremely broad.
Therefore, we adopt the gap edge at the positive bias as $\Delta_1$.

In addition to the energy gap $\Delta_1$, a subgap feature exists at all doping levels, namely, kinks or inflection points appear at $|eV| < \Delta_1$, indicating the existence of another lower energy scale.
The positions of the kinks are indicated by the arrows in Fig.~\ref{SpectrumEvolution} and the inverted triangles in Fig.~\ref{RepresentativeSpectrum}.
This low-energy structure is conspicuous in the spectra with $\Delta_1 > 7$~meV.
As shown in Fig.~\ref{RepresentativeSpectrum}, the existence of the kinks can be also seen in the $\mathrm{d}^3I/\mathrm{d}V^3(V)$ curve; there are two dips (shaded regions) at $|eV| < \Delta_1$, which do not appear in the ideal $d$-wave LDOS.
These kinks symmetrically appear with respect to $E_\mathrm{F}$.
Although the kink energy is somewhat ambiguous due to its broad nature, we estimate the energy from the two minima in the $\mathrm{d}^3I/\mathrm{d}V^3(V)$ curves and hereafter label it as $\Delta_2$.
As can be seen in Fig.~\ref{SpectrumEvolution}, $\Delta_2$ is almost independent of $\Delta_1$ and $x$; the kinks always appear at $V \sim \pm 5$~mV at all doping levels.
This means that $\Delta_2$ is spatially homogeneous and independent of the doping level within our experimental accuracy.
This behavior is in contrast to that of $\Delta_1$.
In Fig.~\ref{GapDistribution}(e), we summarize the doping dependences of $\Delta_1$, $\Delta_2$, and $T_\mathrm{c}$.
With decreasing hole concentration from the overdoped regime, $\Delta_1$ steadily increases while broadening the width of its distribution, whereas $\Delta_2$ is almost constant.
This doping dependence of the $\Delta_1$ distribution is similar to that observed in Bi2212,~\cite{2212_phonon-sts_2006_nature,2212_two-gaps_gomes_2007} and quantitatively agrees well with the antinodal gap in LSCO observed by angle-resolved photoemission spectroscopy (ARPES).~\cite{arpes_lsco_2002}

In this study, the kinks in the tunneling spectra have been observed for La-based cuprates as well as Bi-based cuprates.
As reported in our previous study on La-Bi2201, the kink energy is spatially uniform regardless of the gap energy in LSCO.
Furthermore, our latest STS data on La-Bi2201 for various doping levels also show the independence of the kink energy on doping level.~\cite{2201_machida_pc}
Therefore, these kinks observed in the two distinct materials are probably due to the same entity, and the existence of the two energy scales is thought to be a common characteristic among the cuprate HTSs.
It is well known that the cuprate HTSs have two characteristic regions in momentum space---the nodal and antinodal regions.
Thus, the tunneling spectrum, which involves the electronic states over the whole momentum space, consists of two characteristic energy ranges associated with the two momentum regions; the low-energy part (in the vicinity of $E_\mathrm{F}$) originates from the nodal region, and the high-energy part (near the gap edge) originates from the antinodal region.
Therefore, the two energy scales, $\Delta_1$ and $\Delta_2$, found in this study are associated with the antinodal and nodal electronic states, respectively.
Hence, our results can be interpreted in terms of the nodal and antinodal gaps.
The gap inhomogeneity observed by STS corresponds to the spatial fluctuation of the antinodal gap, and the nodal gap is spatially uniform.
Furthermore, the nodal gap is almost independent of the doping level and $T_{\mathrm{c}}$ throughout the superconducting regime measured, although the antinodal gap steadily increases upon underdoping.
In the case of LSCO ($T_{\mathrm{c}}^{\mathrm{max}}=38$~K), $\Delta_2$ is about 5~meV, while it is about 10~meV in the optimally doped La-Bi2201 ($T_{\mathrm{c}}=34$~K).~\cite{2201_machida_2006}
The energy of the kinks observed by other STS studies are $\sim 25$~meV in underdoped Bi2212 ($T_{\mathrm{c}}=73$~K)~\cite{2212_two-gaps_gomes_2007} and $\sim 7$~meV in (Bi$_{1-y}$Pb$_{y}$)$_{2}$Sr$_{2}$CuO$_{6+\delta}$ ($T_{\mathrm{c}}=15$~K).~\cite{2201_two-gaps_boyer_2007} 
Thus, the lower energy scale, which is thought to be ubiquitous in HTSs, depends on the material rather than on $T_{\mathrm{c}}$.

For the two energy scales observed in the tunneling spectra, there are several experimental reports that give a probable explanation for the appearance of the kink structure inside the energy gap.
Most of the recent ARPES studies have revealed the extremely broad characteristics of the antinodal spectra in contrast to the sharp quasiparticle peak in the nodal spectra,~\cite{arpes_lsco_zhou_2004,arpes_2212_tanaka_2006}
and Fourier transform STS has shown the disappearance of quasiparticle interference near the antinodal region.~\cite{qpi_mcelroy_2003,qpi_hanaguri_2007}
These results suggest the decoherence of the quasiparticles in the antinodal region.
Therefore, the tunneling spectrum will probably show kinks inside the energy gap instead of the simple $d$-wave form by integrating both the coherent nodal quasiparticles with the sharp peak and the incoherent antinodal quasiparticles with the broad nature.
In this case, $\Delta_2$ represents the energy scale where the coherent quasiparticles persist.
Furthermore, the ARPES data of LSCO and La-Bi2201 show substantial deviation of the gap symmetry from the ideal $d$-wave form.~\cite{arpes_lsco_terashima_2007,arpes_2201_kondo_2007,arpes_2nd_gap_lee_2007}
This also indicates the possibility of a certain deformation of the gap shape in the tunneling spectrum.
Moreover, recent ARPES studies claim the coexistence of two distinct energy gaps with different energy scales.~\cite{arpes_2212_tanaka_2006,arpes_2201_kondo_2007,arpes_lsco_terashima_2007,arpes_2nd_gap_lee_2007}
It was shown from the temperature evolution of the energy gap that the antinodal gap and the nodal gap correspond to the pseudogap and the superconducting gap, respectively.
It is feasible to associate the antinodal pseudogap and the nodal superconducting gap with our observed $\Delta_1$ and $\Delta_2$, respectively.
However, even so, the magnitude of the nodal superconducting gap is not in simple agreement with $T_\mathrm{c}$.

\section{Conclusion}
We report on an STM/STS study on superconducting LSCO from the underdoping to the overdoping regime at 4.2 K.
The surface topography of the cleaved $ab$ surface showed one-dimensional modulation, which is consistent with the configuration expected in the LTO phase. 
The tunneling spectra in all samples showed kinks inside the energy gap, indicating the existence of two characteristic energy scales.
Since the kink structure has been observed in both La-based and Bi-based cuprates, the existence of two energy scales is common to the cuprate HTSs. 
The energy of the kink is spatially uniform and almost independent of the doping level, although the energy gap shows spatial inhomogeneity and steadily increases upon underdoping, indicating a difference between the nodal and antinodal states.

It has long been believed that the higher energy scale characterizes the superconductivity.
It is necessary to investigate what the recently discovered lower energy scale characterizes.
Meanwhile, reconsideration of the higher energy scale is also needed.

The present study demonstrates that it is now possible for STM/STS experiments on La-based cuprates to achieve atomic resolution.
This is significant progress in the research of cuprate HTSs and could open the way for the real space investigation of {\it striped} states.~\cite{stripe_tranquada_1995,stripe_review_2003}
Further STM/STS studies on La-based cuprates and their comparison with other materials and experiments (particularly ARPES) will increase the understanding of the electronic states of HTSs.

\end{document}